\documentclass[reprint,
superscriptaddress,
 amsmath,amssymb,
 aps,
]{revtex4-2}

\usepackage{graphicx}
\usepackage{dcolumn}
\usepackage{bm}
\usepackage[colorlinks=true,linkcolor=blue,urlcolor=blue,citecolor=blue]{hyperref}
\usepackage{comment}
\usepackage{xcolor}
\usepackage{amssymb}
\usepackage{lipsum}
\usepackage{physics}

\begin{document}

\preprint{APS/123-QED}

\title{Inferring entropy production in anharmonic Brownian gyrators}

\author{Biswajit Das}
\email{bd18ip005@iiserkol.ac.in}
\affiliation{Department of Physical Sciences, Indian Institute of Science Education and Research Kolkata, Mohanpur Campus, Mohanpur, West Bengal 741246, India}

\author{Sreekanth K Manikandan}
\email{sreekanth.km@fysik.su.se}
\affiliation{Nordita
Stockholm University and KTH Royal Institute of Technology
Hannes Alfvéns väg 12, SE-106 91 Stockholm, Sweden}

\author{Ayan Banerjee}
\email{ayan@iiserkol.ac.in}
\affiliation{Department of Physical Sciences, Indian Institute of Science Education and Research Kolkata, Mohanpur Campus, Mohanpur, West Bengal 741246, India}

\date{\today}

\begin{abstract}
A non-vanishing entropy production rate is one of the defining characteristics of any non-equilibrium system, and several techniques exist to determine this quantity directly from experimental data. The short-time inference scheme, derived from the thermodynamic uncertainty relation, is a recent addition to the list of these techniques. Here we apply this scheme to quantify the entropy production rate in a class of microscopic heat engine models called Brownian gyrators. In particular, we consider models with anharmonic confining potentials. In these cases, the dynamical equations are indelibly non-linear, and the exact dependences of the entropy production rate on the model parameters are unknown. Our results demonstrate that the short-time inference scheme can efficiently determine these dependencies from a moderate amount of trajectory data. Furthermore, the results show that the non-equilibrium properties of the gyrator model with anharmonic confining potentials are considerably different from its harmonic counterpart - especially in set-ups leading to a non-equilibrium dynamics and the resulting gyration patterns.

\end{abstract}

\maketitle


\section{Introduction}
\label{section:1}
Two main characteristics distinguish non-equilibrium systems from their equilibrium counterparts. The first one is the presence of non-vanishing currents in the phase space between (at least) some pairs of states - a manifestation of the breaking of the so-called \textit{detailed-balance} condition ~\cite{gnesotto2018broken}. The second one is the positive rate of total entropy production $\sigma = \frac{\langle \Delta S_{tot}\rangle}{t}$, where $t$ is the time duration of the process ~\cite{seifert2005entropy}. For non-equilibrium systems in a stationary state, $\sigma$ quantifies the rate at which heat is dissipated to the environment, and thus quantifies the thermodynamic cost of maintaining the process ~\cite{seifert2008stochastic}.

There is a considerable amount of literature where detecting phase space currents is used as a \textit{model-independent} means to check whether the system is in equilibrium or not \cite{zia2007probability}. A notable work in this direction is Ref.~\cite{battle2016broken}, which demonstrated that phase-space currents are signatures of the non-equilibrium nature of active fluctuations in microscopic biological systems. Similarly, the non-equilibrium characteristics in the actin cytoskeleton has been quantified by Seara et al. in Ref.~\cite{seara2018entropy}.  In addition, an application to a noise driven linear electric circuit can be found in Ref.~\cite{gonzalez2019experimental}. This approach, however, faces challenges when extended to high dimensional systems, where large amounts of data will be required for the convergence of the (high-dimensional) current estimators. On another note, the entropy production rate $\sigma$ associates a numerical value (in units of $k_B s^{-1}$) to a non-equilibrium system that measures the extent of the non-equilibrium character. Hence there has been a significant amount of interest, mainly within the framework of Stochastic Thermodynamics ~\cite{seifert2012stochastic}, to develop accessible techniques which can quantify the entropy production rate from a moderate amount of phase-space trajectory data ~\cite{li2019quantifying,manikandan2020inferring,frishman2020learning,otsubo2020estimating,van2020entropy}. A significant recent addition to this list is the thermodynamic uncertainty relation \cite{barato2015thermodynamic}, which demonstrated that a lower bound to the entropy production rate could be obtained in
terms of the fluctuations of arbitrary currents $J$ in the phase space, as has been described in Ref.~\cite{li2019quantifying}, 

\begin{align}
\label{eq:zero}
    \sigma \geq \left[\frac{2 k_B \langle J \rangle^2}{t\;\text{Var}(J)}\right].
\end{align}
The average and the variance are computed over an ensemble of currents of length $t$, which can be straightforwardly constructed from the phase-space trajectories of the system. More recently, for a large class of non-equilibrium systems with a continuous-space and continuous-time dynamics, it was shown that this inequality saturates in the short time limit, and an exact estimate of the entropy production rate can be obtained from  \cite{manikandan2020inferring,otsubo2020estimating,van2020entropy},
\begin{align}
\label{eq:one}
    \sigma = \lim_{\Delta t \rightarrow 0}\max_J \left[\frac{2 k_B \langle J_{\Delta t} \rangle^2}{\Delta t\;\text{Var}(J_{\Delta t})}\right]
\end{align}
Here $J_{\Delta t}$  is an arbitrary current of length $\Delta t$   constructed from the trajectory data sampled at an interval $\Delta t$. The average current and variance are computed over an ensemble of currents of length $\Delta t$ measured from a single stationary trajectory. Furthermore, the optimal current $J^*$ which maximizes the RHS of Eq.\ \eqref{eq:one} is known to be proportional to $\Delta S_{tot}$, and helps determine the thermodynamic force field  $\textbf{\textit{F}}(\textbf{x})$ in the phase-space \cite{manikandan2020inferring,otsubo2020estimating,van2020entropy}. In recent work,  the feasibility of this inference scheme was tested on experiments with a colloidal particle in a time-varying potential \cite{manikandan2021quantitative}, where it was shown that the method correctly reproduces the entropy production rate and the thermodynamic force field  known by other theoretical means. 

\begin{figure}[ht]
     \centering
     \includegraphics[scale =0.25]{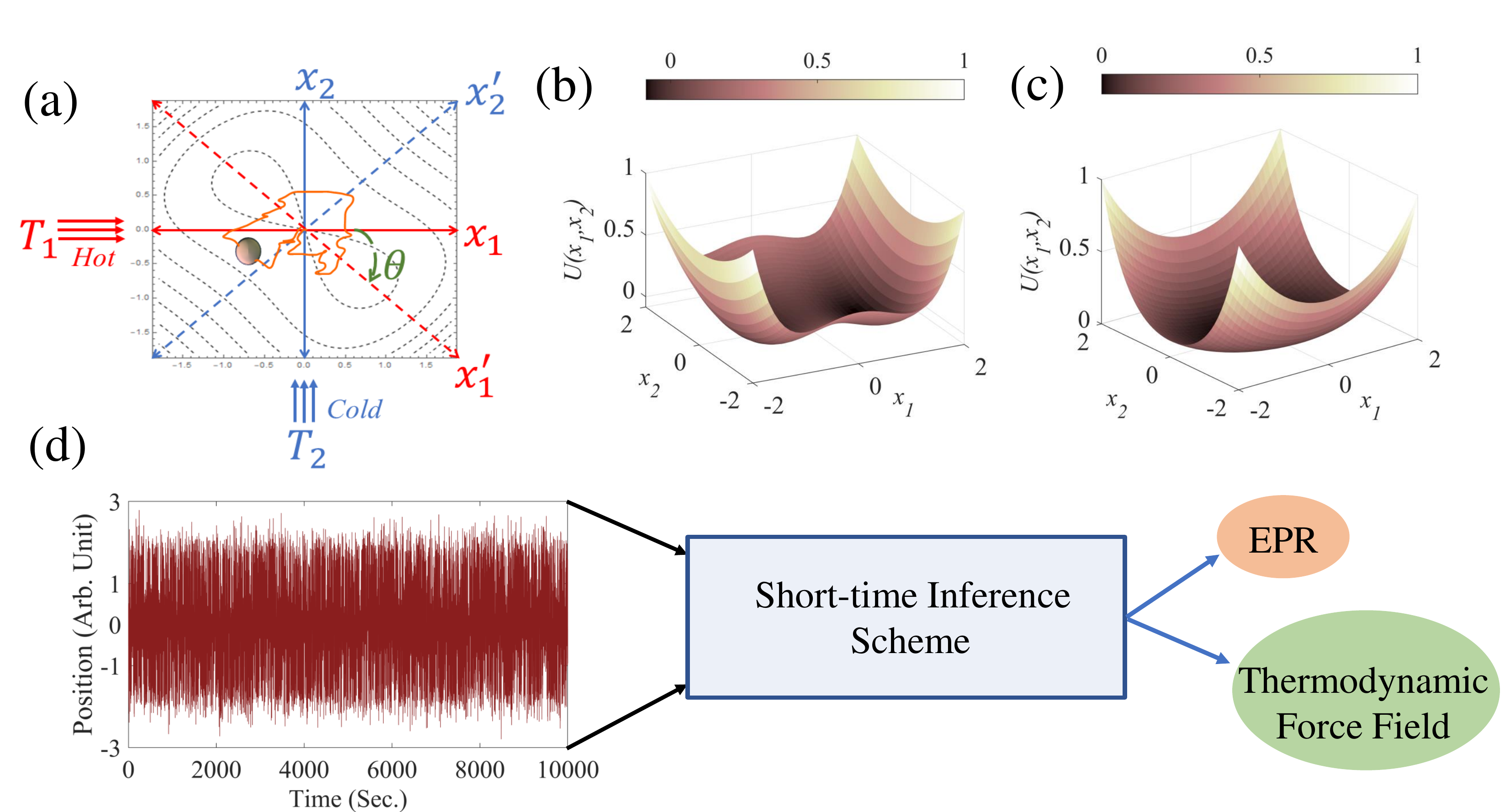}
    \caption{\textbf{Brownian gyrator with complex potential:}  (a) Brownian gyrator model: A Brownian particle is confined in a generic potential in the presence of two heat baths of different temperatures ($T_1$   and $T_2)$ along two axis. In some cases, the potential needs to be rotated by an angle $\theta$ along with $T_1 \neq T_2$ to obtain the gyration effect of the trapped particle. Shapes of the  generic potentials of our study: (b) Double well potential  and (c) Quartic potential. (d) Short-time inference technique is used in this study to characterise the non-equilibrium features of the system through the estimation of entropy generation rate and the thermodynamic force field from the trajectory of the trapped particle.}
     \label{fig:model}
 \end{figure}

Here we apply Eq.~\eqref{eq:one} to quantify the entropy production rate and to obtain the thermodynamic force field in an interesting class of microscopic non-equilibrium systems called autonomous Brownian gyrators~ \cite{filliger2007brownian,argun2017experimental,chang2021autonomous}. They consist of a micrometer scale colloidal particle in a confining potential, coupled to two thermal reservoirs in the orthogonal directions. When the temperatures of the reservoirs are different, and for certain forms of the confining potential, the dynamics of the particle break the \textit{detailed-balance} condition, making the system non-equilibrium. However, the exact dependencies of the phase space currents or the entropy production rate on the parameters of the gyrator are only known when the confining potential is a quadratic one \cite{filliger2007brownian, argun2017experimental,manikandan2019efficiency}. In this work, we consider Brownian gyrators in anharmonic confining potentials \cite{chang2021autonomous} and demonstrate that Eq.\ \eqref{eq:one} can be used to determine how the entropy production rate depends on the model parameters. Furthermore, we obtain the non-trivial thermodynamic force fields and velocity fields in the phase space,  whose characteristics are considerably different from their harmonic counterparts.

The paper is organized as follows. In section \ref{section:2} , we reproduce the previously known closed-form expressions for the entropy production rate as well as the thermodynamic force field for the Brownian gyrator in a harmonic potential well. In Section \ref{section:3}, we briefly describe the short-time inference scheme. In section \ref{section:4}, we apply this scheme to two examples of anharmonic Brownian gyrators ~\cite{chang2021autonomous}, to determine how the entropy production rate depends on the system parameters. We also obtain the thermodynamic force field and the phase space velocity fields in different cases, and discuss their qualitative features. In all cases, we also discuss how the results compare with the corresponding ones for the Brownian gyrator in a quadratic potential. In Section \ref{section:5}, we conclude with an outlook towards experimental demonstrations and future work.

\section{Theoretical Model}
\label{section:2}
The dynamics of the Brownian gyrator in the overdamped limit can be described by the Langevin equations: 
\begin{equation}
     \gamma_1 \dot{x}_{1} = -\frac{\partial U (x_1, x_2)}{\partial x_1} + \sqrt{2 \gamma_1 k_B T_1} \xi_1(t).
     \label{eq:lan_2d1}
\end{equation}
\begin{equation}
    \gamma_2 \dot{x}_{2} = -\frac{\partial U (x_1, x_2)}{\partial x_2} + \sqrt{2 \gamma_2 k_B T_2} \xi_2(t).
    \label{eq:lan_2d2}
\end{equation}
Here $U(x_1,x_2)$ is the 2D confining potential in the $x_1$, $x_2$ plane. $x_1$ and $x_2$ are further coupled to two different thermal reservoirs at temperatures $T_1$ and $T_2$ respectively. The corresponding thermal noises are denoted by $\xi_i(t)$, which are both Gaussian noises with $\langle \xi_i(t) \rangle = 0 $ and  $\langle \xi_i(t) \xi_j(t')\rangle = \delta_{ij}\delta(t-t') $. Here $\gamma_i$ is the viscous drag coefficient of the medium, which is related to the temperature of the medium through the Einstein relation $D_i\gamma_i = k_BT_i$, where $k_B$ is the Boltzmann constant.  For simplicity, in our case we keep $\gamma_1=\gamma_2 = \gamma$ and $k_B=1$.

\indent   Brownian gyrators in quadratic confining potentials are well-studied both theoretically and experimentally \cite{argun2017experimental,chiang2017electrical,dotsenko2013two,filliger2007brownian}. In this case, the confining potential has the form:
\begin{equation}
    U_{har}(x_1, x_2) = \frac{1}{2} (x_1 \ x_2)\cdot \mathbf{R(-\theta)}\cdot \mathbf{k}\cdot \mathbf{R(\theta)}\cdot \begin{pmatrix} x_1 \\ x_2 \end{pmatrix}
    \label{eq:harmonic_pot}
\end{equation}
where $\mathbf{R(\theta)}$ is the $2D$ rotation matrix corresponding to the  angle of rotation $\theta$,
\begin{equation}
\mathbf{R(\theta)} = \begin{pmatrix}
\cos \theta & -\sin \theta \\
\sin \theta & \cos \theta
\end{pmatrix}
\end{equation}
Note that such quadratic potentials are naturally occurring in the case of optical traps or tweezers. The matrix $\mathbf{k}$ determines the stiffness of the trapping potential in orthogonal directions, and has the form:
\begin{align}
    \mathbf{k} = \begin{pmatrix}
k_{1} & 0 \\
0 & k_{2} 
\end{pmatrix}.
\end{align}
The resulting dynamics of the system can be described using a linear diffusion equation of the form
\begin{align}
    \mathbf{\dot{x}} = -\mathbf{A}  \;\mathbf{x} + \mathbf{B}\;\boldsymbol{\xi}
\end{align}
where $\mathbf{x} = (x_1,\;x_2)$, and
\begin{equation}
    \mathbf{A} =\begin{pmatrix}
   {\frac{k_{1}}{\gamma}}\cos^{2}\theta + {\frac{k_{2}}{\gamma}}\sin^{2}\theta  &   ({\frac{k_2}{\gamma}} - {\frac{k_1}{\gamma}})\sin\theta \cos\theta \\
    ({\frac{k_2}{\gamma}} - {\frac{k_1}{\gamma}})\sin\theta \cos\theta & {\frac{k_{1}}{\gamma}}\sin^{2}\theta + {\frac{k_{2}}{\gamma}}\cos^{2}\theta
    \end{pmatrix}
\end{equation}
and,
\begin{equation}  
\begin{split}
    \mathbf{B} &= \begin{pmatrix}
    \sqrt{2k_{B}T_1/\gamma} & 0 \\
    0 &\sqrt{2k_{B}T_2/\gamma}
    \end{pmatrix}.
\end{split}
\end{equation}
The probability of finding the trapped particle at a position \textbf{x} at time $t$ can be determined in terms of the probability density function $\rho(\textbf{x},t)$, which obeys a Fokker-Plank equation
\begin{equation}
\begin{split}
    \partial_t \rho(\mathbf{x},t) &= - \nabla \cdot (-\mathbf{A}\mathbf{x}\rho({\mathbf{x},t}) - \mathbf{D} \nabla \rho(\mathbf{x},t)) \\
    & \equiv -\nabla \cdot \mathbf{J}(\mathbf{x},t)
    \end{split}
    \label{eq:fp}
\end{equation}
with $\mathbf{D} = \mathbf{BB}^T/2$. Here $\mathbf{J}(\mathbf{x},t)$ is the probability current in the phase space.\\

\begin{figure}
    \centering
    \includegraphics[width=\linewidth]{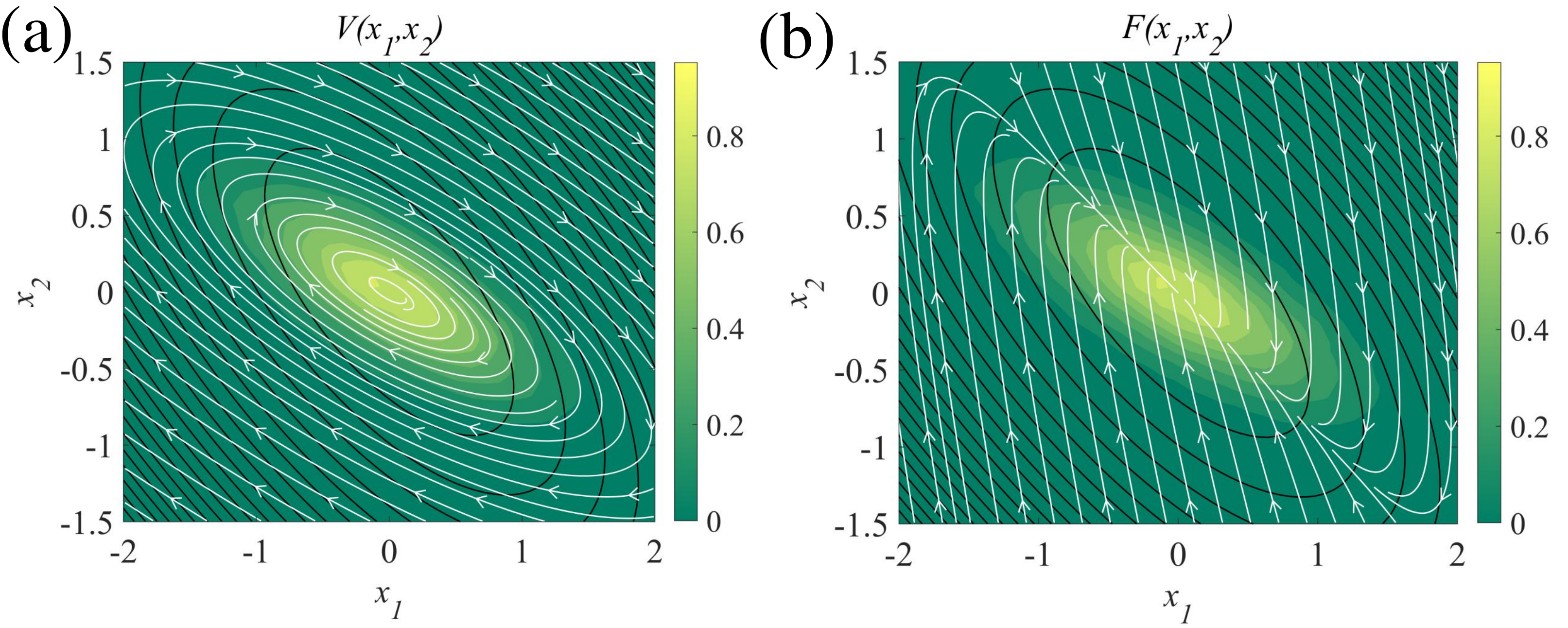}
    \caption{(a) Velocity field and, (b) thermodynamic force field of Brownian gyrator with harmonic confining potential are plotted (streamlines) on top of the steady-state probability (color maps) of the position of the particle. Parameters used: $k_1 = 1$, $k_2 = 5$ and, $\theta = 45^\circ$. Ratio of the temperatures along the orthogonal axis is fixed at, $\alpha = {\frac{T_2}{T_1}} =  0.1$. The closed loops denote the equipotential contours.}
    \label{fig:harmonic}
\end{figure}

In the $t\rightarrow \infty$ limit, the system can be shown to reach a non-equilibrium stationary state with a characteristic distribution and current given by,
\begin{align}
\begin{split}
    \rho_{ss}(\mathbf{x}) &= (2\pi \sqrt{\det \mathbf{ C}})^{-1} e^{-\frac{1}{2}\mathbf{x}^T\mathbf{C}^{-1}\mathbf{x}} \\
    \mathbf{J}_{ss}(\mathbf{x}) &=  (-\mathbf{A}\mathbf{x} + \mathbf{D}\mathbf{C}^{-1}\mathbf{x})\rho_{ss}(\mathbf{x}),
    \end{split}
    \label{eq:rho_current}
\end{align}
where the co-variance matrix $\mathbf{C}$ can be written as ~\cite{argun2017experimental},

\begin{widetext}
\begin{equation}
    \mathbf{C} = \frac{1}{\Tr  \mathbf{A} \ \det  \mathbf{A}}\begin{pmatrix}
    D_{2}A_{12}^{2} + D_{1}(A_{22}^2 + \det  \mathbf{A}) &  -D_{1}A_{21}A_{22} - D_{2}A_{11}A_{12} \\
    -D_{1}A_{21}A_{22} - D_{2}A_{11}A_{12} & D_{1}A_{21}^{2} + D_{2}(A_{11}^2 + \det  \mathbf{A})
    \end{pmatrix}
    \label{eq:har_covar}
\end{equation}
\end{widetext}
In Fig.\ \ref{fig:harmonic}(a), we plot $\rho_{ss}(\mathbf{x})$ and the phase-space velocity field $\mathbf{V}(\mathbf{x})=\frac{\mathbf{J}_{ss}(\mathbf{x})}{\rho_{ss}(\mathbf{x})} $ for a particular choice of parameters. It can be noticed that the velocity field faithfully follows the probability density contours.

 Using standard definitions in stochastic thermodynamics, the entropy production rate in the steady state can then be obtained as the integral ~\cite{lander2012noninvasive, seifert2005entropy}
\begin{equation}
\begin{split}
   \sigma &= \int d\mathbf{x}\; \mathbf{F}(\mathbf{x})\cdot \mathbf{J}_{ss}(\mathbf{x}),
    \label{eq:sigma_brgr_harmonic}
    \end{split}
\end{equation}
where $\mathbf{F}(\mathbf{x})$ is the local conjugate thermodynamic force field associated with the steady-state current, defined as $\mathbf{F}(\mathbf{x}) = k_{B} \mathbf{J}^{T}_{ss}(\mathbf{x})\cdot {\mathbf{D}}^{-1}/\rho_{ss}(\mathbf{x})$. In Fig. \ref{fig:harmonic}(b), we show the thermodynamic force field for the parameter choice in Fig. \ref{fig:harmonic}(a).

Since we can individually determine all the terms in the integrand of Eq.\ \eqref{eq:sigma_brgr_harmonic}, it can be explicitly evaluated, giving,
\begin{align}
\label{eq:sigmaQuadratic}
     \sigma =k_B \frac{(k_{1} - k_{2})^2  (T_1 -  T_2)^2 \sin^2 (2 \theta)}{4(k_{1} + k_{2}) T_1 T_2\gamma}
\end{align}
This equation shows that when $k_1=k_2$ (corresponding to an isotropic harmonic potential) or $T_1=T_2$, or when $\theta = \frac{n\pi}{2}$, the entropy production rate vanishes, and the system will remain in an equilibrium state. For other combinations of parameter values, it's straightforward to notice that $\sigma > 0$, consistent with the Second Law of thermodynamics.

Note that in a generic setting, where the underlying confining potential is of degree $> 2$, or in other words - the diffusion matrix $\mathbf{D}$ is dependent on $\mathbf{x}$ - a closed form expression for the integrand in Eq.~\eqref{eq:sigma_brgr_harmonic} is not available. Thus, if we consider the case of anharmonic Brownian gyrators, exact expressions of the entropy production rate are also not known. In the following, we describe how these challenges can be overcome using the short-time inference scheme.
\begin{figure*}
     \centering
     \includegraphics[width=0.99\textwidth]{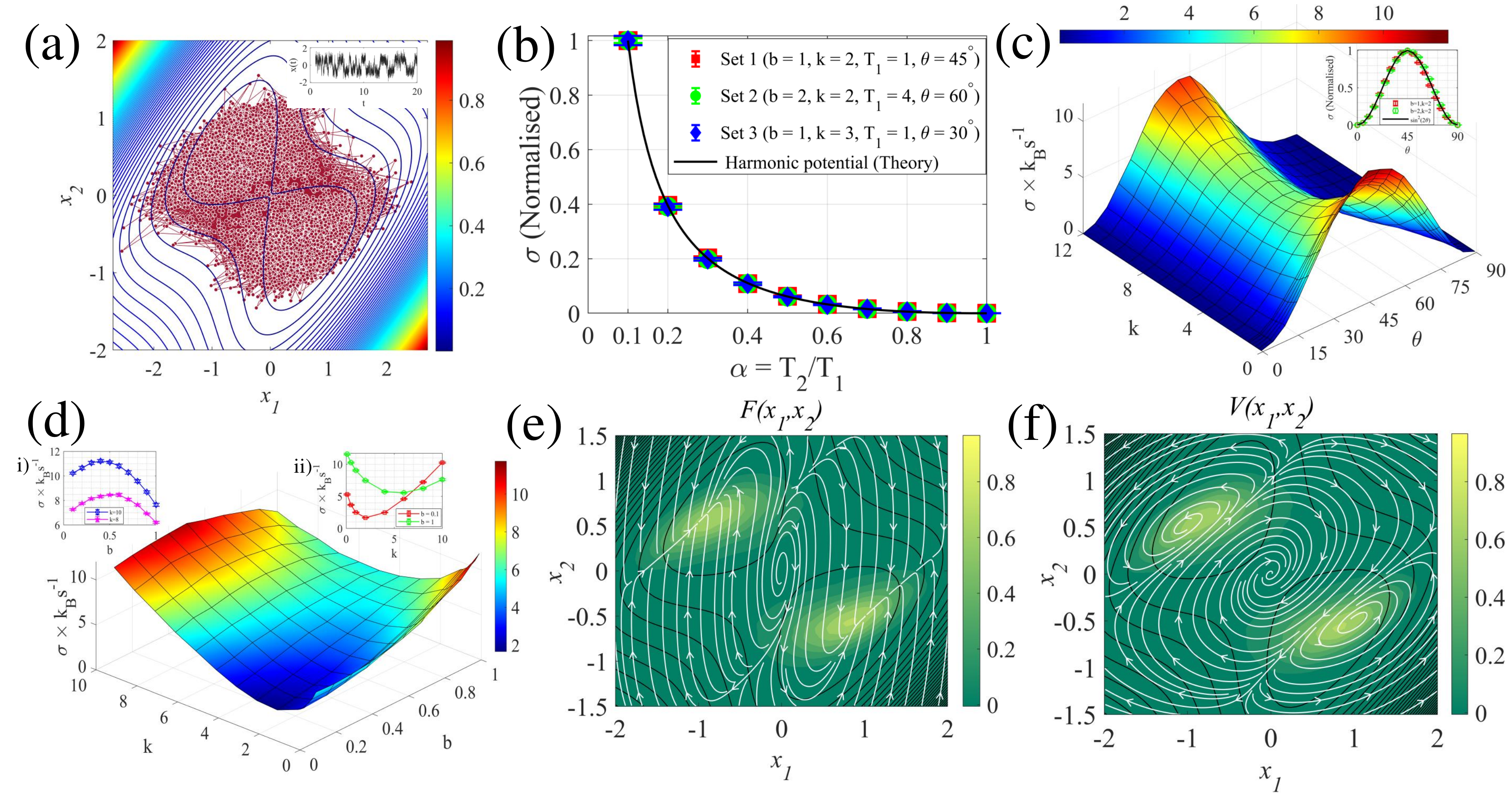}
     \caption{\textbf{Quantitative analysis of $\sigma$ corresponding to the Brownian Gyrator model with the bi-stable potential:} (a) Particle trajectories in the 2D co-ordinate space of the bi-stable potential when the potential is rotated by an angle $\theta = 45^\circ$. Parameter values: $b =1,k = 2, \alpha =\frac{T_2}{T_1} = 0.1$. The displacement of the particle shows jumps along any direction due to the presence of finite barrier height of the potential (inset). (b) Entropy production rate ($\sigma$) of the system is quantified as a function of the ratio of temperatures ($\alpha = T_2/T_1$) along the two axes in different conditions of the gyrator indicated by the different values of the parameters mentioned in the legend. $\sigma$ of each sets are normalised by dividing with the $\sigma$ value corresponding to $\alpha = 0.1$ of that particular set. (c) Absolute value of $\sigma$ is plotted as a function of angle of rotation ($\theta$) and the stiffness constant ($k$) of the harmonic part of the potential while keeping the value of $\alpha$ = 0.1. (Inset: $\sigma$ is plotted with $\theta$ at fixed $k$, and shows $\sim \sin^2(2\theta)$ behaviour.) (d) $\sigma$ is quantified as a function of the parameters `$k$' and `$b$' with $\alpha = 0.1$ and $\theta = 45^\circ$.  (e) Thermodynamic force field and (f) the phase space velocity field plotted as streamlines. In (e) - (f), the color maps represent the steady state probability distribution of the system, while the black loops denote the equipotential contours.  Error bars are given as the standard deviation over 10 independent measurements of the entropy production rate $\sigma$ for a fixed choices of model parameters.}
     \label{fig:tur_dw}
 \end{figure*}

\section{The short-time inference scheme}
\label{section:3}

In this paper, we have used the recently discovered short-time inference scheme to estimate the entropy production rate and the thermodynamic force field from the time-series data. This method was first introduced in Ref.\ \cite{manikandan2020inferring}, rigorously proved in Refs.~\cite{otsubo2020estimating,van2020entropy}, and recently tested in a colloidal experimental setup in Ref.~\cite{manikandan2021quantitative}. Using this technique, we can obtain the steady-state entropy production rate as,
\begin{equation}
    \sigma = {\lim_{\Delta t \rightarrow 0}}\max_{J} \Big[ \frac{2 k_B \langle J \rangle ^2}{\Delta t \text{Var}(J)}\Big],
    \label{eq:unc}
\end{equation}
where $J$ is a weighted scalar current defined as
\begin{align}
    J = \mathbf{d}\Big( \frac{\mathbf{x}^{i+1} + \mathbf{x}^{i}}{2}\Big)\cdot(\mathbf{x}^{i+1} - \mathbf{x}^{i})
    \label{eq:discrete_current}
\end{align}
where, $\mathbf{x}^i$ denotes the $d-$ dimensional time discretised trajectory data  in a time interval $\Delta t \ll \lbrace \tau_s \rbrace$, where $\lbrace \tau_s \rbrace$ is the set of all relevant time scales in the system. The superscript $i$ denotes the discrete time labels and $\langle \cdot \rangle$ denotes the ensemble average. 
In principle, $\mathbf{d}(\mathbf{x})$ can be any arbitrary  $d-$ dimensional function, characterized by an infinite number of degrees of freedom that correspond to the function's value at each point $\mathbf{x}$. However, as usual in inference problems, we have a finite stochastic trajectory at hand, from which only a finite amount of information can be inferred \cite{frishman2020learning}. It is therefore natural to approximate the $\mathbf{d}(\mathbf{x})$ as as the linear combination of a finite set of basis functions
  $\boldsymbol{\psi}_m(\mathbf{x})$ as,
    \begin{align}
        \mathbf{d(x)} = \sum_{m=1}^{M}\text{diag}\left( \mathbf{C}_m \right)\cdot\boldsymbol{\psi}_m(\mathbf{x}).
        \label{eq:dx}
    \end{align}
    where, $\mathbf{C}_m =\left[c_m^1,\;c_m^2,\; ..., \; c_m^d\right] \in \mathbb{R}^{d}$, the notation diag$(\mathbf{C}_m)$ corresponds to a diagonal matrix whose entries are elements of the vector $\mathbf{C}_m$,  and $\boldsymbol{\psi}_m(\mathbf{x}) = \left[\psi_m^1(\mathbf{x}),\;\psi_m^2(\mathbf{x}),\; ..., \; \psi_m^d(\mathbf{x})\right]$.
    
    The basis functions then define a set of $ N = d\times M $ number of basis currents $\lbrace\phi_{n=(k-1)\times M + m} \rbrace= \lbrace\int_0^{\Delta t}  \psi_m^k(\mathbf{x})\circ dx_k \rbrace$, where $m = 1,\;2,\;...,\; M$ and $k = 1,\;2,\;...,\; d$ and the symbol $\circ$ denotes the Stratanovich convention for the stochastic integral. 
Then the maximization problem in Eq.\ \eqref{eq:unc} over the space of currents straightforwardly translates to a maximization problem over the space of $N$ number of coefficients $\lbrace c_m^k \rbrace$. Interestingly, for any fixed choice of the basis functions, the analytical solution to this optimization problem is known  \cite{van2020entropy}, and is given by

\begin{align}
    \sigma = \frac{2 \langle \phi_r \rangle (\boldsymbol{\Xi}^{-1})_{r,s} \langle \phi_s \rangle }{\Delta t}
    \label{eq:sigma_analytical}
\end{align}
where $\boldsymbol{\Xi}$ is an $N\times N$ dimensional correlation matrix, whose elements are $(\boldsymbol{\Xi})_{r,s} = \langle \phi_r \phi_s \rangle - \langle \phi_r  \rangle \langle  \phi_s \rangle$. The
optimal coefficients can be obtained as,
\begin{align}
    c^{k*}_m  &=\frac{(\boldsymbol{\Xi}^{-1})_{r,s} \langle \phi_s \rangle}{[\langle  \phi_r \rangle (\boldsymbol{\Xi}^{-1})_{r,s} \langle  \phi_s \rangle ]},& r&=(k-1)\times M+m.
    \label{eq:optimal_coeff}
\end{align}
In both Eq.\ \eqref{eq:sigma_analytical} and Eq.\ \eqref{eq:optimal_coeff}, the Einstein summation convention is assumed and repeated indices are summed over.
For the derivation of the proof, please refer to section IIB of Ref.\ \cite{van2020entropy}.
The corresponding optimal force field is given by $\mathbf{d}^* \equiv \sum_{m=1}^{M}\text{diag}\left(\mathbf{C}_m^*\right)\cdot \boldsymbol{\psi}_m(\mathbf{x})$, where $\mathbf{C}_m^* =\left[c_m^{1*},\;c_m^{2*},\; ..., \; c_m^{d*}\right]^T$.
Furthermore, the thermodynamic force field  $  \mathbf{F}$ is known to be proportional to $\mathbf{d}^*$, ($\mathbf{F} = \nu\; \mathbf{d}^*$) and the proportionality constant can be determined as $\nu = \frac{\text{Var} (J^*)}{2 \langle J^* \rangle}$  \cite{manikandan2020inferring,otsubo2020estimating}. This follows from the fact that for overdamped diffusive processes, $\frac{\text{Var}(\Delta S_{tot})}{\langle \Delta S_{tot} \rangle } \rightarrow 2$ as $t\rightarrow 0$. See Eqs. (3) - (4) Ref.\ \cite{manikandan2020inferring} and Eqs. (20) - (29)  in Ref.\ \cite{otsubo2020estimating} for proofs. 

    We remark that the basis functions $\boldsymbol{\psi}_m(\mathbf{x})$ can be chosen from any complete basis such as polynomials, wavelets or Fourier modes \cite{frishman2020learning}. Due to the completeness property, it is clear that a large enough set of these functions will span the vector space. In this work, we use the polynomial basis in two dimensions. As we show later, it can be verified that a third order polynomial is sufficient for inference for the examples we look at. 

 \section{Results and Discussions}
 \label{section:4}
 \begin{figure*}
     \centering
     \includegraphics[width=0.99\textwidth]{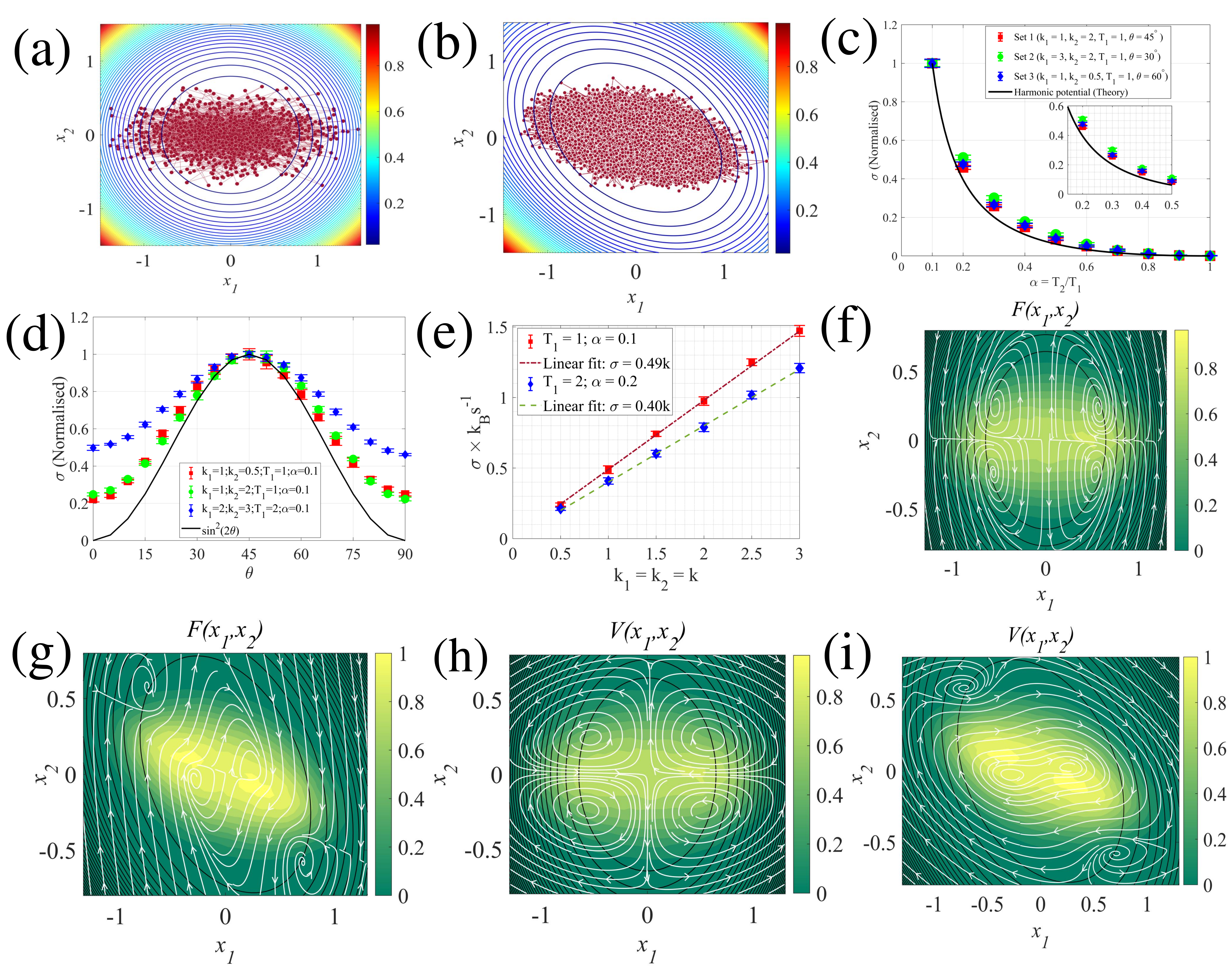}
     \caption{\textbf{Quantitative analysis of $\sigma$ corresponding to the Brownian Gyrator model with the quartic potential:} (a) Brownian trajectories of the trapped particle in the isotropic quartic potential with $k_1 = k_2 = 1$ and $\alpha = 0.1$. (b) Stochastic trajectories of the trapped particle in the anisotropic quartic potential with $k_1 =1 , k_2 = 2$ and $\alpha = 0.1$. (c) Entropy production rate ($\sigma$) of the system with the anisotropic quartic potential is quantified as a function of the ratio of temperatures ($\alpha = T_2/T_1$) along two axis in different conditions of the gyrator indicated by the different values of the parameters mentioned in the legend. $\sigma$ of each sets are normalised by dividing with the $\sigma$ value corresponding to $\alpha = 0.1$ of that particular set. (d) $\sigma$ is plotted as a function of $\theta$ for the anisotropic potential with $k_1 = 1, k_2 = 2$. We observe deviation from $\sim sin^{2}(2\theta)$ as $\sigma \neq 0$ for $\theta = 0$. (e) Absolute value of $\sigma$ shows a linear nature with the parameter $k$ for the gyrator system with isotropic quartic potential. (f) The thermodynamic force field of the system with the isotropic quartic potential ($k_1 = k_2 = 1 , \alpha = 0.1$). (g) The thermodynamic force field of the system in the anisotropic quartic potential ($k_1 = 1, k_2 = 2, \alpha = 0.1$). (h) Velocity field for the system with isotropic quartic potential ($k_1 = k_2 = 1 , \alpha = 0.1$). (i)  Velocity field for the system with the anisotropic quartic potential ($k_1 = 1, k_2 = 2, \alpha = 0.1$). In (f) - (i) color maps represent the steady state probability distribution of the system while the black loops denote the equipotential contours.  Error bars are given as the standard deviation over 10 independent measurements of the entropy production rate $\sigma$ for a fixed choices of model parameters.}
     \label{fig:tur_qu}
 \end{figure*}

In this section, we apply the short-time inference scheme to the Brownian gyrator in anharmonic potentials. The gyration characteristics of these systems were studied in great detail in Ref.~\cite{chang2021autonomous} using numerical simulations and analyses based on the Fokker-Planck equation. They observed that the gyrating patterns in case of anharmonic confining potential are significantly distinct from the equiprobable contour lines of the potential. In contrast, steady-state currents for the harmonic case faithfully follow the tangent of the equiprobability contour. The paper also discussed the positivity of the steady-state entropy production rate, but no quantitative characterization, such as how it depends on the system parameters, was discussed. Here we address this issue using the short-time inference scheme. In relevant cases, we also compare the results with the findings in \cite{chang2021autonomous}.

We first consider the double well potential \cite{chang2021autonomous} (Fig.\ref{fig:model}(b)) given by
\begin{equation}
    U_{bs}(x_1^\prime,x_2^\prime) = {x_1^\prime}^4 -{2}b{x_1^\prime}^2 + \frac{1}{2}k{x_2^\prime}^2
    \label{eq:bs_potential}
\end{equation}
where $x_1^\prime$ and $x_2^\prime$ are two axis of the potential which is rotated by an angle $\theta$ with respect to the axis of the temperatures ($x_1,x_2$) as,  
\begin{equation}
    \begin{pmatrix}
    x_1^\prime \\ x_2^\prime 
    \end{pmatrix}
    = \begin{pmatrix}
    cos\theta  &  -sin\theta\\ sin\theta & cos\theta
    \end{pmatrix}
    \times\begin{pmatrix}
    x_1 \\ x_2
    \end{pmatrix}
    \label{eq:rot}
\end{equation}

The parameter $b$ can be used to control the nature of the bi-stable part of the potential along $x_1^\prime$, as the position of the minima ($\sim\pm \sqrt{b}$) and the barrier height ($\sim b^2$) of the potential are dependent on it. The harmonic part of the potential along $x_2^\prime$ is characterised by the stiffness constant $k$ along that direction. \\

There are two natural timescales in this problem. The first one is the relaxation timescale in the harmonic part of the trap, and is given by $\tau_\gamma = \gamma/k$. The second one is the inverse of the Kramers escape rate, which we have determined numerically in all cases. See Appendix \ref{appendix:A} for details. Further, to apply the short-time inference scheme, we first generate the stationary trajectories of the system using first order Euler integration, which we proceed to apply to Eq.\eqref{eq:lan_2d1} and Eq.\eqref{eq:lan_2d2} with a time step of $\Delta t = 0.001 s$, which is chosen such that it is at least one order of magnitude less than the two relevant timescales for all the parameter choices. As initial conditions, we choose $x_0, y_0$ from a Gaussian distribution with mean $[0,0]$ and standard deviation $[\sqrt{D_1},\sqrt{D_2}])$. We then run the simulation for a certain time ($\sim 10000  s$ ) which is several orders of magnitude higher than the relevant timescales in the problem, so that the system reaches its steady state distribution unambiguously. The subsequent time-series data of length $10000  s$ is used for applying the short-time inference scheme.

To apply the short-time inference scheme, we choose a polynomial basis of order 3. The corresponding basis currents are given by 
\begin{align}
\begin{split}
\phi_1&= \int_0^{\Delta t} 1\circ dx_1,\\
    \phi_2&= \int_0^{\Delta t} x_1\circ dx_1,\\
    \phi_3&= \int_0^{\Delta t} x_2\circ dx_1,\\
    \phi_4&= \int_0^{\Delta t} x_1^2\circ dx_1,\\
    ...\\
    \phi_{10}&=\int_0^{\Delta t} x_2^3\circ dx_1,\\
    \phi_{11}&= \int_0^{\Delta t} 1\circ dx_2,\\
    \phi_{12}&= \int_0^{\Delta t} x_1\circ dx_2,\\
    \phi_{13}&= \int_0^{\Delta t} x_2\circ dx_2,\\
    \phi_{14}&= \int_0^{\Delta t} x_1^2\circ dx_2,\\
    ...\\
    \phi_{20}&=\int_0^{\Delta t} x_2^3\circ dx_2.
    \end{split}
     \label{eq:basis_current}
\end{align}
The notation $\circ$ stands for the Stratanovich convention, where the integral is evaluated as in Eq.\ \eqref{eq:discrete_current}.
Using these basis currents and Eq.\ \eqref{eq:sigma_analytical}, we can determine the  entropy production rate $\sigma$ for a fixed set of parameters. By changing the parameters and repeating the same steps, the dependence of $\sigma$ on the various parameters of the system can be determined. Here we follow this procedure, and first determine the dependence of $\sigma$ of the parameters $\alpha (= \frac{T_2}{T_1})$ and $\theta$. 

If the confining potential is harmonic, we know from Eq.\eqref{eq:sigmaQuadratic}, that $\sigma \propto \frac{(T_1-T_2)^2}{T_1T_2} \sin^2(2\theta)$. From the analysis, we find that $\sigma$ has exactly the same dependencies on these parameters in case of the double well. In Fig.\ \ref{fig:tur_dw}(b), we show the dependence of $\sigma$ on $\alpha$ for different values of $k$, $b$,  $\theta$ and $T_1$. The plots are normalized by the value at $\alpha = 0.1$. We see that they fall on top of each other, and agree with the functional behaviour given in Eq.\ \eqref{eq:sigmaQuadratic}. In Fig.\ \ref{fig:tur_dw}(c), we plot $\sigma$ as a function of $\theta$ (and $k$), which shows the $\sigma \propto \sin^2(2\theta)$ behaviour for a fixed $k$ (see the inset).  The highest value of entropy production is obtained when when $\theta = \pi/4$ for all $k$.  We also find that the entropy estimate is close to $0$ in the cases $\alpha = 1$ or $\theta = n\pi/2$. This is expected, since these limits corresponds to the equilibrium limits of the dynamics (When $\alpha = 1$, the system is in  equilibrium with the reservoirs at temperature $T_1=T_2=T$. When $\theta = n\pi/2$, the two degrees of freedom are decoupled, and are independently in equilibrium with the two reservoirs at temperatures $T_1$ and $T_2$.). 

We also find that $\sigma$ depends non-monotonically on the parameters $b$ and $k$. In Fig.\ \ref{fig:tur_dw}(c), we show that $\sigma$ is minimized for a particular value of $k$ for any value of $\theta$. Similarly, $\sigma$ has both minima and maxima  in the $b - k$ space as shown in Fig.\ \ref{fig:tur_dw}(d). It is tempting to interpret this observation as the display of a resonance-like behaviour, where certain configurations are able to maximally (minimally) produce entropy by exploiting the spatially anisotropic temperature gradient. However, the physical origins of this behaviour remains elusive to us presently, and merits deeper investigation. It is also clear that this may have interesting applications in experiments - especially in the design and optimization of  microscopic engines.

Using the short-time inference scheme, we can further obtain the conjugate thermodynamic force field $\mathbf{F}(\mathbf{x})$ using Eq.\ \eqref{eq:optimal_coeff}. It  characterizes the spatial dependence of the entropy production rate, and can be also used to compute the total entropy production ($\Delta S_{tot}$) along a single stationary trajectory as, 
\begin{align}
    \Delta S_{tot} (t) = \int_{x(0)}^{x(t)} \mathbf{F}(\mathbf{x})\circ d\mathbf{x}.
\end{align}
In Fig.\ref{fig:tur_dw}(e), we plot $\mathbf{F}(\mathbf{x})$ for a particular choice of parameters. Furthermore, when the matrix $\mathbf{D}$ is known (as in our case), it is possible to obtain the average phase space velocity field ($\mathbf{V}(\mathbf{x})$) as $\mathbf{V}(\mathbf{x}) = \mathbf{D} \cdot \mathbf{F}(\mathbf{x})$. This is shown in Fig.\ref{fig:tur_dw}(f). 
Notably, the velocity field lines do not circulate about the equiprobability contours of the particle. This feature is in agreement with the findings of Chang et al. about the gyrating characteristics of probability currents \cite{chang2021autonomous}. \\

As the second example, we consider a Brownian gyrator model with a quartic potential given by
 \begin{equation}
    U_{qu}(x_1^\prime,x_2^\prime) = (k_1{x_1^\prime}^ 2+ k_2{x_2^\prime}^2)^2
    \label{eq:qu_potential}
\end{equation}
where $x_{1}^\prime$ and $x_{2}^\prime$ are the axis of the potential rotated by an angle $\theta$ with respect to the co-ordinate frame ($x_1$,$x_2$). The potential can be categorised as isotropic ($k_1 = k_2$) and anisotropic ($k_1 \neq k_2)$ depending on the values of the stiffness constants ($k_1$ and $k_2$) along the axis of the potential. 
Relaxation timescales corresponding to the stiffness constants are the natural timescales ($\tau_i = \gamma/k_i \ , i=\{1,2\}$) for this potential. To apply the short-time inference scheme, the stationary trajectories for this confining potential are generated in the exact same way as discussed for the previous potential, using the Euler integration scheme with a time step $\Delta t = 0.001 \ s$ which is at least one order of magnitude less than $\tau_i$ .
\begin{figure*}
    \centering
    \includegraphics[width=0.99\textwidth]{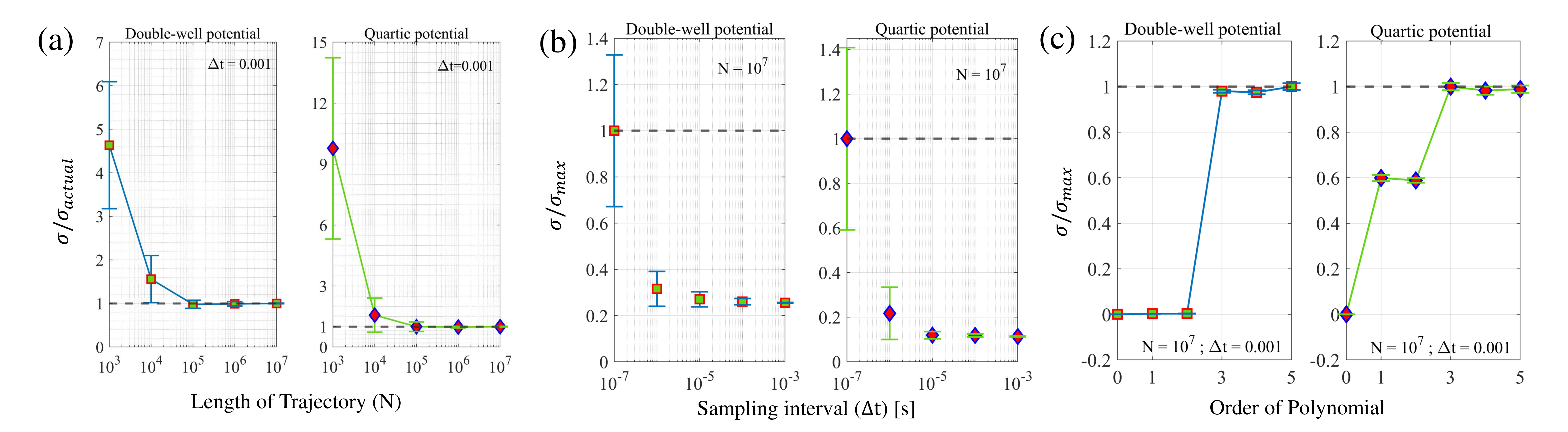}
    \caption{\textbf{Effect of hyperparameters on the inference of Entropy production rate ($\sigma$): } $\sigma$ is plotted as a function of  (a) length of the trajectory, (b) sampling interval, and (c) order of the polynomial basis for both double-well and quartic confining potential.  Error bars are given as the standard deviation over 10 independent measurements of entropy production rate $\sigma$ for a fixed choices of model parameters. [\textbf{Parameters:} \textbf{Double-well potential:} $b = 1, k = 2, \theta = 45^\circ, \alpha = 0.1$, \textbf{Quartic well potential:} $k_1 = 1, k_2 = 2, \theta = 45^\circ, \alpha = 0.1$]}
    \label{fig:sigma_order}
\end{figure*}

The steady state entropy production rate for different parameter values can be estimated just as in the previous case. In Fig.\ \ref{fig:tur_qu}(c), we show the dependence of $\sigma$ with $\alpha$ for different values of $k_1$, $k_2$ and $\theta$, which is found to be slightly deviating from the line corresponding to the $\sigma$ of the gyrator with harmonic confining potential. Indeed, even the error bars are smaller compared to the deviation from the behaviour anticipated in case of the harmonic confining potential. We believe further investigation with analytical rigour is required to confirm whether the dependence of entropy production rate on the two temperatures is universal across different potential wells. On the other hand, the dependence of $\sigma$ on $\theta$ is found to significantly deviate from the previous cases ($\sim sin^{2}(2\theta)$) as depicted in Fig.\ \ref{fig:tur_qu}(d). In particular, $\sigma$ is non-zero even when $\theta = 0^\circ$. This is because, in a quartic potential, the motion of the particle along the two directions are coupled even when $\theta = 0$. For the same reason, as opposed to the Brownian gyrator in an isotropic ($k_1 = k_2 = k$) quadratic potential, the entropy production in an isotropic quartic potential will be non-zero. For the isotropic potential, $\sigma$ is estimated to be  monotonically increasing with the stiffness constant ($k$) as shown in Fig.\ \ref{fig:tur_qu}(e). 

 In Fig.\ \ref{fig:tur_qu}(f) and  Fig.\ \ref{fig:tur_qu}(g) we plot $\mathbf{F}(\mathbf{x})$ for the isotropic and the anisotropic potential, respectively. Furthermore, the average phase-space velocity field ($\mathbf{V}(\mathbf{x})$) profiles for these potentials are also obtained and shown in Fig.\ \ref{fig:tur_qu}(h) [isotropic case] and  Fig.\ \ref{fig:tur_qu}(i) [anisotropic case]. For the isotropic case, four circulating regions are revealed in the thermodynamic force field and also in the velocity field. Just as we found in the previous case, the field lines do not follow the equiprobable contours. This is again in agreement with the findings in Ref.\ \cite{chang2021autonomous}.

Finally, we look at the effect of the hyper-parameters of the inference problem. These are the length of the trajectory $N$, sampling interval ($\Delta t$), and the order of the polynomial used for the inference scheme. In
 Fig.\ref{fig:sigma_order}(a), we show the dependence of the inferred entropy production rate on the length of the trajectory used for inference. The results show that a trajectory with $10^4 -10^5$  points with sampling interval $\Delta t = 0.001\ s$ will be sufficient to give a reliable estimate of the entropy production rate. Indeed, with more data, the accuracy of the estimate is found to be better, with less statistical error. In
 Fig.\ref{fig:sigma_order}(b), we demonstrate the dependence of the inferred value of the entropy production rate on the sampling interval  $\Delta t$ for a fixed number of points in the trajectory. Theoretically, it is known that the inference scheme is dependent on $\Delta t$ and gives the closest estimate to the actual entropy production rate when we take the $\Delta t \rightarrow 0$ limit. Our results show that this is indeed the case, and smaller $\Delta t$ values lead to an increase in the estimated value of the entropy production rate. However, if we keep the number of points in the trajectory fixed, as in
 Fig.\ref{fig:sigma_order}(b), we find that smaller $\Delta t$ values lead to higher statistical errors.  
 
 In Fig.\ref{fig:sigma_order}(c), we look at the effect of the order of the polynomial used for the inference of entropy production rate. For both the anharmonic gyrator models, the order of the polynomial that appear in the drift term of the Langevin equation is three. For the double-well case, we find that no significant entropy is inferred for order $< 3$. Interestingly, this implies that the harmonic contribution to the entropy production rate is absent in this case. For the quartic well, we find that a non-zero value of entropy production is inferred already when an order one polynomial is used, indicating that the harmonic contribution to entropy production exists. Again, the inference is seen to saturate at order three. We remark that in a generic case, where we do not know the actual degree of the non-linearity of the problem, an iterative procedure of this kind can be used to obtain a reliable and close-to-true-value estimate of the entropy production rate. Similarly, we can also fix the number of basis functions required to accurately represent the optimal, thermodynamic force field.
 
\section{Conclusions}
\label{section:5}
In summary, we have demonstrated that the short-time inference scheme (Eq.\ \eqref{eq:one}) can be used to characterize the non-equilibrium character of Brownian gyrators with complex potential energy landscapes quantitatively. We considered a double-well and a quartic potential in two dimensions and determined how the entropy production rate depended on the parameters of the potentials as well as the temperatures along orthogonal directions. For specific parameter choices, we also obtain the thermodynamic force field and the phase-space velocity field, without resorting to statistical binning techniques.

Our results suggests that the exact dependence of entropy production rate on the two temperatures is apparently universal across gyrators with different confining potentials. However,  further theoretical investigations are required to substantiate whether this is indeed the case. For the Brownian gyrator in a  double-well potential, we find that the entropy production rate non-monotonically depends on the parameter which controls the bi-stable nature of the potential. We also notice that the contribution to entropy production entirely comes from the cubic nonlinearity of the driving forces in the system since a linear or second order basis used for inference captures no significant entropy generation from the trajectories.  In the case of the Brownian gyrator with a quartic confining potential, this is not the case, and we find that a significant contribution to entropy production comes from a first order truncation of the basis functions.  Such considerations will be crucial for practical applications of the inference scheme to complex non-equilibrium systems such as biological systems, where anharmonic energy landscapes  naturally arise \cite{angeli2004detection,sali1994does,frauenfelder1991energy,marcucci2013attached}.

It will be interesting to observe if our results can be tested in experimental realizations of anharmonic Brownian gyrators, which can be set up using higher-order, structured Gaussian beams~\cite{chai2012improvement,liu2019enhancement} or with feedback optical tweezers systems ~\cite{kumar2018nanoscale}. We hope to explore these aspects in future work.

\section{Acknowledgements}
Biswajit Das is thankful to Ministry Of Education of Government of India for the financial support through the Prime Minister's Research Fellowship (PMRF) grant. Nordita is partially supported by Nordforsk.

\appendix

\section{Determining the Kramers rates}
\label{appendix:A}
For symmetric double well potentials, the Kramers escape rate corresponds to the rate at which transitions take place from one well to the other. Except in a limited number of cases, the exact analytic dependence of the Kramers escape rate on the parameters of the model is unknown. See \cite{dunkel2003kramers} for a comprehensive analytic treatment of the 1D Kramers problem for particle in a one dimensional double well potential, in contact with a thermal reservoir at temperature $T$. In contrast, the gyrator setup we consider in this work is much harder to be treated analytically, as it is a 2-dimensional system which is in contact with two thermal reservoirs  along orthogonal directions. However, it is much more straight forward to track this problem numerically. To this end, we closely follow the discussion in Section III of \cite{dunkel2003kramers}.

\begin{figure}
    \centering
    \includegraphics[width=\linewidth]{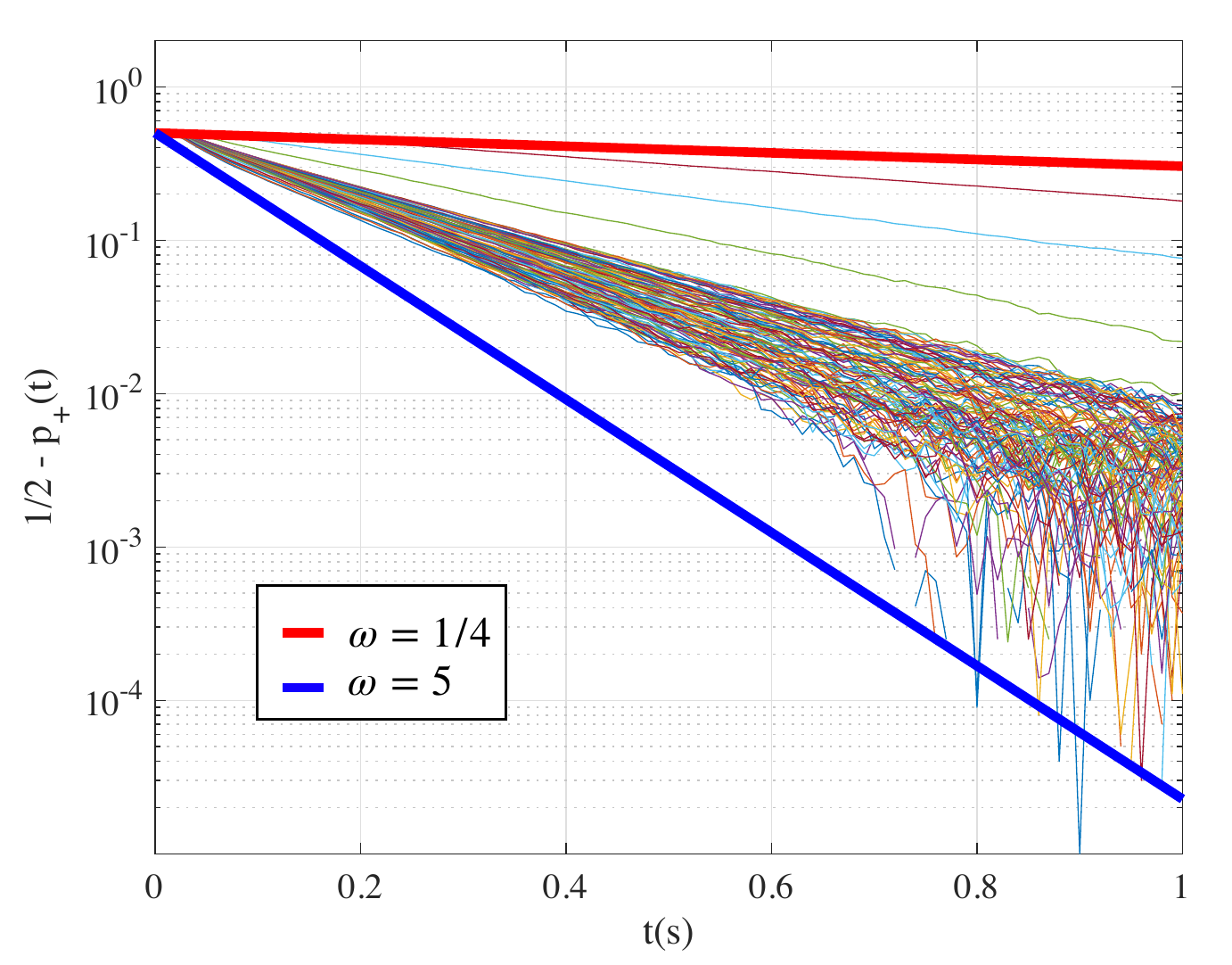}
    \caption{\textbf{Kramers rates for double-well potential:} The quantity $1/2 - p_{+}(t)$\cite{dunkel2003kramers} is plotted for all configurations of the double-well potential we study.}
    \label{fig:Kramers}
\end{figure}

We first classify all the states with $y> \cot (\theta) x$ to be in state $\vert + \rangle$ and the rest to be in state $\vert - \rangle$. Then we initialize an ensemble of particles at the point $(x,y)=(-\cos(\theta),\sin(\theta))$, $\theta \in \left[ 0,\pi/2\right]$. The quantity we intend to estimate analytically is the escape
rate $\omega$ characterizing the decline of the population in the left
well ($p_-(t)$), or the increase of the population in the right well ($p_+(t)$). Due to the symmetry of the potential well, the rate of transition from state $\vert + \rangle$ to state $\vert - \rangle$ will be the same as the rate of transition from state $\vert - \rangle$ to state $\vert + \rangle$. We can therefore follow the exact same derivation as given in  \cite{dunkel2003kramers} and obtain,
\begin{align}
\label{soln}
    p_\pm(t)= \frac{1}{2}\mp\frac{1}{2}\exp(-2\omega t)
\end{align}

In Fig.\ \ref{fig:Kramers}, we plot $1/2-p_+(t)$ for all the parameters we studied (thin lines). We find that the curves are well bounded by curves of the form Eq.\ \eqref{soln} with $\omega =1/4\ s^{-1}$ and $\omega =5 \ s^{-1}$. This means the minimum Kramers time scale for our parameter choices is $>1/5=0.2 s$, and the maximum Kramers time scale is $<4 s$. 

The other relevent time scale in the problem is the relaxation timescale in the optical trap, which is given by $\tau_\gamma = \frac{\gamma}{k}$, where $k$ is the stiffness of the trap. For the particular parameters we have chosen, we have $0.08s\leq\tau_{\gamma}\leq 1s$.
Considering these, for our numerical simulations, We have chosen a time step $\Delta t=0.001s$, which is an order of magnitude less than all the relevant timescales. Similarly, after initializing the system in an arbitrary Gaussian distribution, we discard trajectory data of length $10000s$, several orders of magnitude higher than the relevant timescales, to make sure that the analysis is performed on a stationary time series. Afterwards, we continue the simulation for another $\sim 10000s$ and used this trajectory to infer $\sigma$.

\providecommand{\noopsort}[1]{}\providecommand{\singleletter}[1]{#1}%

\end{document}